\documentclass[letterpaper,11pt]{article}
\usepackage{amsmath, amsthm, amssymb}    
\usepackage{bridges}			
\usepackage{graphicx}			
\usepackage[colorlinks=true, urlcolor=blue, citecolor=black, linkcolor=black]{hyperref}  
\usepackage{subcaption}			

\title{Classical Optics for Charged Black Holes}

\author{Steve Trettel
\vspace{10pt}\\
University of San Francisco, California, USA; strettel@usfca.edu} 

\date{}					

\begin{document}

\maketitle

\thispagestyle{empty}

\vspace{-5mm}

\begin{abstract}
We describe a method for rendering multiple extremally charged black holes using an analogous system in classical optics, simplifying the prerequisite mathematics for generating accurate images. A primary goal of this work is to showcase a suite of techniques from geometry and relativity that may be of interest to illustrators and artists.
\end{abstract}

\vspace{5mm}

\begin{figure}[h!tbp]
\centering
\includegraphics[width=\textwidth]{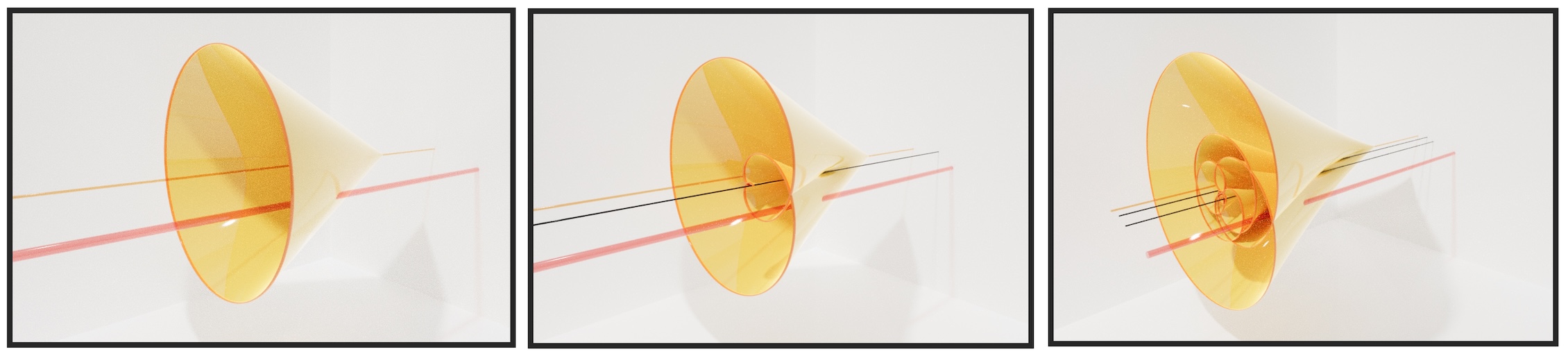}
\caption{Past lightcones for an observer (yellow worldline) raytracing a stationary object (red world-tube) in Minkowski space (left), near a single black hole (center) and near two black holes (right).}
\label{fig:lightcones}
\end{figure}

\noindent Accurately ray-tracing a scene that contains several black holes is, in principle, straightforward: follow every lightlike geodesic backward along the camera's past light cone until it strikes an object (Figure \ref{fig:lightcones}). In practice however, the task explodes into solving nonlinear ODEs on a complicated spacetime (itself the solution of nonlinear PDE), an enterprise that is both fragile and computationally demanding.


\vspace{-2mm}

In this note we exhibit a situation where these difficulties can be avoided, thanks to a beautiful \emph{exact} solution to general relativity containing charged black holes in static equilibrium. Exploiting its symmetries with tricks of Lorentzian geometry, we reduce the null-geodesic equation to a three-dimensional system corresponding to light rays in flat space with variable optical media. This allows the exact rendering of black holes in a \emph{Euclidean} rendering engine with advanced optics - no relativistic simulator required. As examples, all figures were produced with my custom renderer, by implementing variable indices of refraction.

\vspace{-2mm}

\begin{figure}[h!tbp]
\centering
\includegraphics[width=\textwidth]{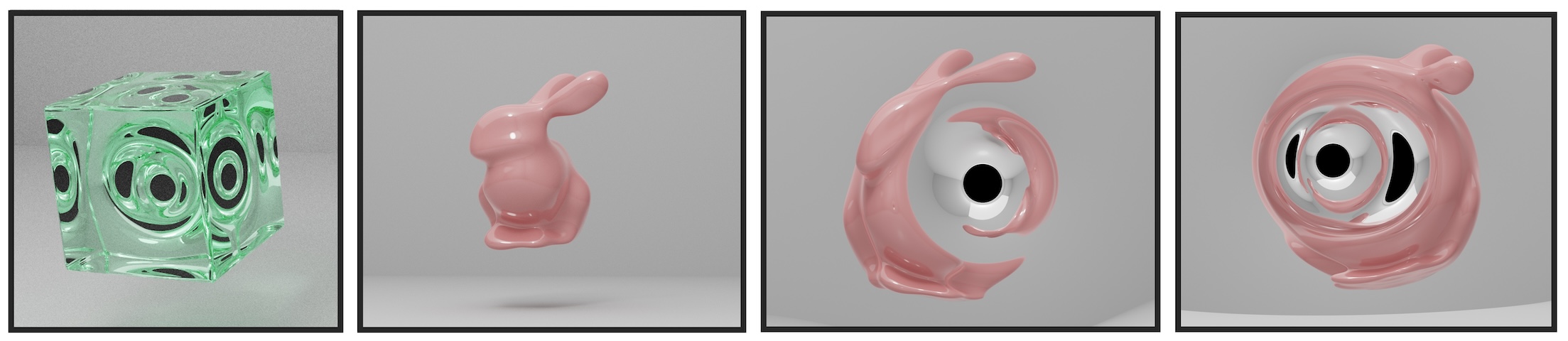}
\caption{A block of glass whose classical optics corresponds exactly to the null geodesics around a binary system of extremal charged black holes (far left).  Flat space raytracing in an optical medium equivalent to Minkowski space (left), a single black hole (center) and two black holes (right). }
\label{fig:bunnies}
\end{figure}

\section*{A Standoff between Electromagnetism and Gravity }

Einstein’s field equation is a system of 10 coupled, nonlinear partial differential equations governing the interplay between spacetime geometry and matter.  Solutions describing single black holes have been known since the theory's inception, but finding multiple black hole solutions is far more challenging as gravitational attraction leads to rapid orbits, energy loss via gravitational waves, and eventual mergers. The strong nonlinearities of Einstein’s equations typically make such spacetimes difficult even with advanced numerics.

Happily there is a beautiful exception to this general rule.  In 1947, Mamjudar and Papapetrou \cite{majumdar} discovered a remarkably simple \emph{exact} solution to the vacuum Einstein Maxwell equations: given a harmonic function $U$ on $\mathbb{R}^3$, the metric $g=-U^{-2}dt^2+U^2ds^2$ 
is a solution\footnote{Throughout, we write $ds^2$ for the Euclidean line element $ds^2=dx^2+dy^2+dz^2$.} for the electrostatic potential $U^{-1}$. Later in 1972 Hartle and Hawking showed these solutions include descriptions of multiple charged black holes \cite{HartleHawking}, henceforth \emph{MP Black Holes}.
Precisely, the potential $U=1+1/r$ represents a single extremally charged black hole (with unit mass and charge, in natural units), and linear combinations $U=1+\sum_i m_i/|r-p_i|$ represent a collection of extremal black holes of mass=charge $m_i$ at coordinate positions $p_i$\footnote{The points $p_i$ actually represent the \emph{event horizons}: following Hartle \& Hawking, one must change coordinates and analytically continue to represent the black hole interiors. But for the tracing light to an external observer, these coordinates are sufficient.}.  This solution is remarkably simple: its \emph{static}, so the presence of multiple black holes does not lead to gravitational waves or mergers.  The secret lies in a precise balancing act, with gravitational attraction \emph{exactly canceled} by electrostatic repulsion.

\vspace{-3mm}
\section*{When Gravity Acts Like Glass}

Starting from this exact solution, we apply results from Lorentzian geometry to simplify the calculations required for tracing light rays. Our key trick is that for computing null geodesics (paths which travel at the speed of light), its often possible to swap out the given spacetime with a  simpler one. Formally:

\noindent {\bfseries Proposition: } \emph{Null geodesics are invariant under a conformal rescaling of the metric.}

\noindent We sketch the proof for the interested reader's benefit: let $\gamma$ be a null geodesic of $g$, and $\tilde{g}=e^{2f}g$ a conformal rescaling.  Note $\gamma$ is still a null curve for $\tilde{g}$ as $\tilde{g}(\dot{\gamma},\dot{\gamma})=e^{2f} g(\dot{\gamma},\dot{\gamma})=0$. Computing $\tilde{\nabla}_{\dot{\gamma}}{\dot{\gamma}}=2df(\dot{\gamma})\dot{\gamma}$ we see its \emph{not turning} (acceleration is parallel to velocity), but is not constant speed.  Changing speed via reparameterization $\tilde{\gamma}=\gamma\circ\varphi$ for $\varphi\colon\mathbb{R}\to\mathbb{R}$, we see
(after some algebra) that $\tilde{\nabla}_{\dot{\tilde{\gamma}}}\dot{\tilde{\gamma}}=\varphi^\prime \left(\varphi^{\prime\prime}+2\varphi^{\prime} df(\dot{\gamma}\circ\varphi)\right)(\dot{\gamma}\circ\varphi)$, and so taking $\varphi$ to solve the differential equation $\varphi^{\prime\prime}+2\varphi^{\prime} df(\dot{\gamma}\circ\varphi)=0$ makes $\tilde{\gamma}$ a null geodesic.  But $\tilde{\gamma}$ traces the same curve in spacetime as $\gamma$,  QED.

\noindent {\bfseries Consequence: } \emph{ The trajectories of light around the extremal black holes described by $g=-U^{-2}dt^2+U^2 ds^2$ are identical to the null geodesics of $\tilde{g}=-dt^2+U^4 ds^2$ (the result of a conformal rescaling by $U^2$).}

This new metric is much simpler: $t$ represents proper time for all stationary observers.  Time's uniform flow renders it inessential to the dynamics, and we lose no information projecting onto their spatial shadows.

\begin{figure}[h!tbp]
\centering
\includegraphics[width=\textwidth]{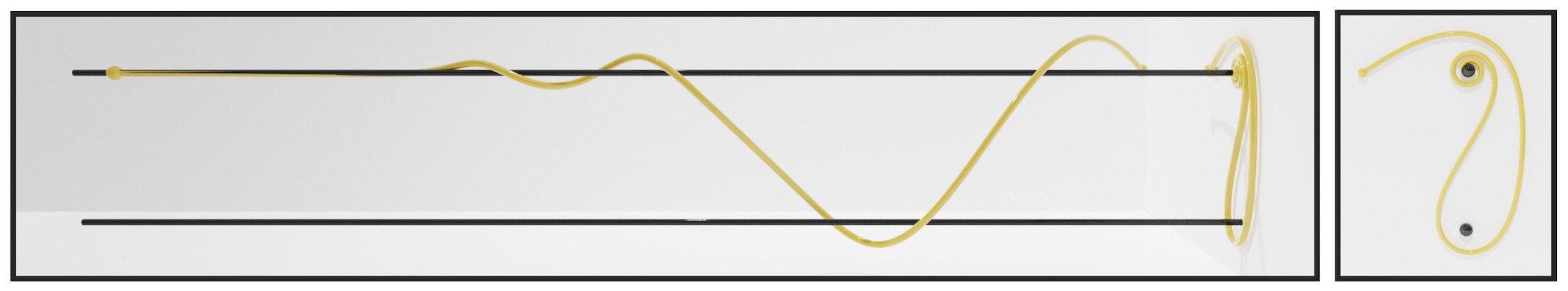}
\caption{Spacetime diagram of a binary black hole with null geodesic (left) and its spatial shadow (right).}
\label{fig:projection}
\end{figure}

\vspace{-3mm}

\noindent {\bfseries Proposition:}
\emph{Null geodesics of $g=-dt^2+g$ on $\mathbb{R}^4$ project to Riemannian geodesics of $(\mathbb{R}^3,g)$.}

\noindent {\bfseries Consequence:} \emph{Let $S\subset \mathbb{R}^3$.  Then the light ray of the MP black hole metric originating at $(p_0,t_0)$ in direction $v_0$ hits the world-tube $\mathbb{R}\times S$ if and only if the geodesic of $(\mathbb{R}^3,U^4 ds^2)$ with initial data $(p_0,v_0)$ intersects $S$.}

Thus for \emph{static objects} (whose worldlines are also unchanging in $t$), this reduces the entire relativistic raytracing problem to three dimensions.
And a final beautiful insight remains hidden in the particular form of this three dimensional system.  Re-interpreting the \emph{distance} computed via arclength integral as \emph{time}, we invoke Fermat's principle to recognize these as \emph{light trajectories in classical physics}.

\noindent{\bfseries Proposition:} 
\emph{Geodesics of the Riemannian metric $n^2 ds^2$ on $\mathbb{R}^3$ are precisely the trajectories of light in classical physics passing through a medium with index of refraction $n$ in Euclidean space $(\mathbb{R}^3,ds^2)$.}

\noindent Putting everything together, we have proven the following rather remarkable correspondence:

\noindent {\bfseries Theorem:} \emph{Raytracing static objects in the MP black hole system with metric $g=-U^{-2}dt^2+U^2(dx^2+dy^2+dz^2)$ is equivalent to tracing objects in flat space, immersed in a medium of refractive index $n=U^2$.}

Note this is a special property of MP black holes: applying the same tricks to the Schwarzschild metric projects light to the geodesics of $g=dr^2/(1+1/r)^2+r^2/(1+1/r)d\Omega^2$ on $\mathbb{R}^3$ (in spherical coordinates), but this metric is \emph{not} conformal to the flat metric $dr^2+r^2d\Omega^2$, so these are not the light rays of a classical system.

\begin{figure}[h!tbp]
\centering
\includegraphics[width=\textwidth]{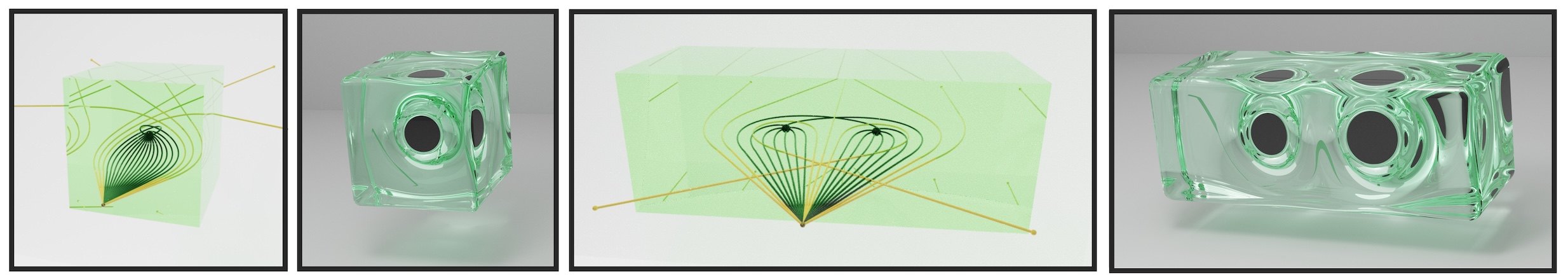}
\caption{The geodesics of $U^4 ds^2$ correspond to the null geodesics of MP black holes.  Inside a material with refractive index $U^2$ light follows these geodesics\textemdash an optical analog of these black holes. }
\label{fig:glass}
\end{figure}

\vspace{-6mm}

\section*{Building a Renderer}



\begin{figure}[h!tbp]
\centering
\includegraphics[width=\textwidth]{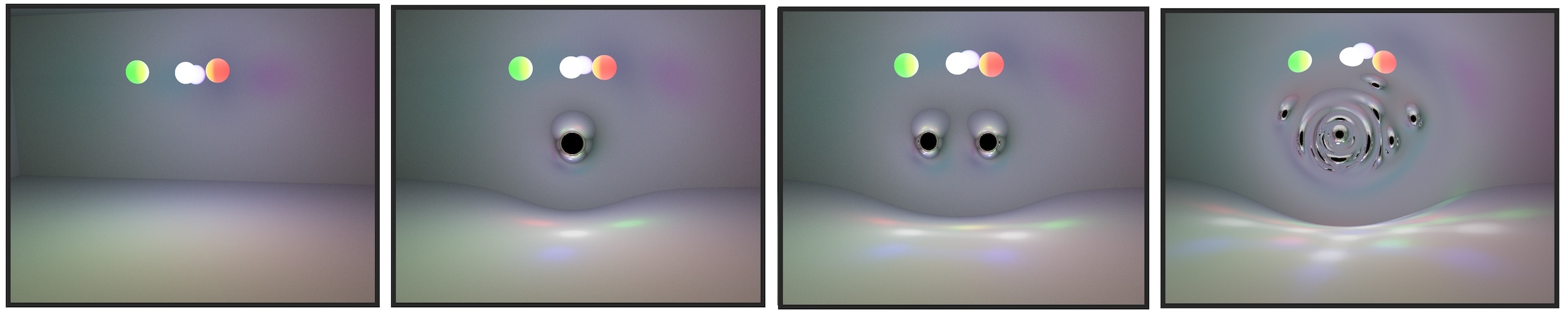}
\caption{The lensing of four lights (white, red, green and blue) in a room with Minkowski geometry (far left), one black hole (center left), two black holes (center right) and nine black holes (far right). }
\label{fig:lensing}
\end{figure}

As a result, any renderer capable of supporting spatially varying refractive indices can accurately depict charged black holes, by filling a region with optical medium of index $U^2 = \left(1+\sum_i m_i/|r-p_i|\right)^2$: see Fig. \ref{fig:glass} for my own implementation. You can also fully convert a Euclidean renderer to a MP black hole renderer (as I did for Figs. \ref{fig:bunnies} and \ref{fig:lensing}). First replace affine rays with numerically integrated geodesics of the metric $U^4\,ds^2$.  Then replace the tracing / raymarching loop
with a simple check for object intersections at each integration step.
And that's it! As $U^4\,ds^2$ is conformally flat, its inner product is a multiple of the standard inner product. So normals, reflections and refractions—anything using normalized dot products — carry over unchanged.  For those wanting to modify a hobby renderer, we give the precise ODEs for the geodesic flow below.

\noindent {\bfseries Proposition:} The trajectories $\gamma(t)=(x(t),y(t),z(t))$ of light in an optical medium with index of refraction $n=U^2$ solve the differential equation $U\ddot{\gamma}=2\|\dot{\gamma}\|^2\nabla U - 4(\dot{\gamma}\cdot \nabla U)\dot{\gamma}$.

\noindent After this hard work we treat ourselves to an image of \emph{gravitational lensing}: like more familiar caustics in path tracing, convergence requires many samples, but the result is stunning for multi-black hole spacetimes.


\section*{Black Hole Shadows}

\begin{figure}[h!tbp]
\centering
\includegraphics[width=\textwidth]{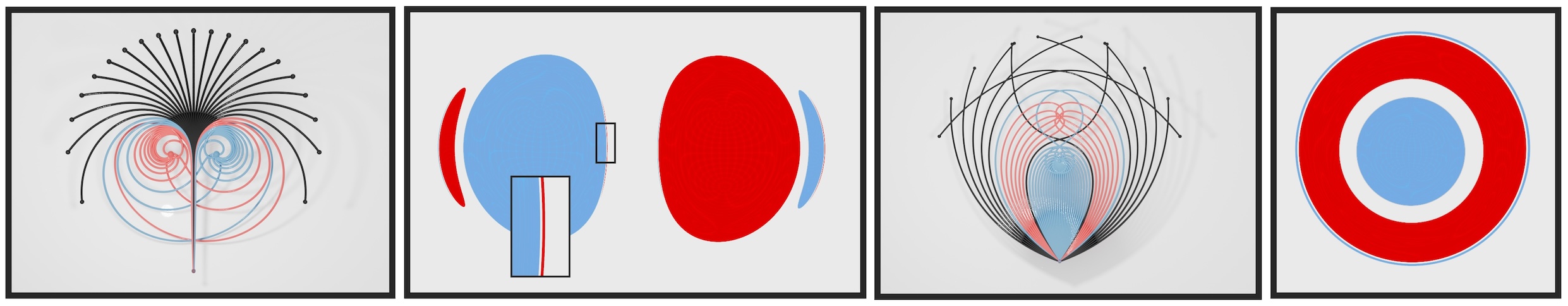}
\caption{Perspectives on a binary black hole system, showing geodesics to the viewer (left), and what they see (right). Horizon-penetrating rays and the corresponding shadows are (artificially) colored. }
\label{fig:chaos-binary}
\end{figure}

\begin{figure}[h!tbp]
\centering
\includegraphics[width=\textwidth]{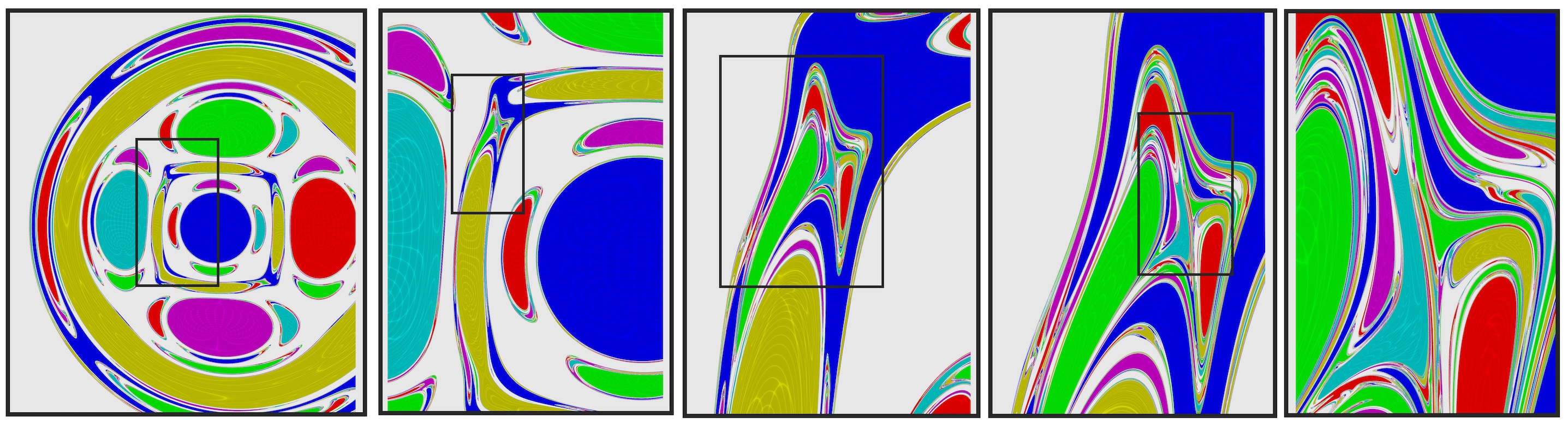}
\caption{A system of six MP black holes at the vertices of an octahedron from a distant observer’s view at multiple levels of zoom, revealing the chaotic trajectories.  Each horizon is given a distinct color.
}
\label{fig:chaos-zoom}
\end{figure}

These techniques also allow real-time rendering of black hole \emph{shadows} (or event horizons, as viewed by a distant observer).  There is a collection of papers on the chaotic dynamics of light in these MP black hole spacetimes (see \cite{Kluitenberg} for a readable introduction) whose results we can illustrate, by tracing light rays forward from the screen and coloring pixels according to which black hole they fall into. Figure \ref{fig:chaos-binary} gives two perspectives on a binary system, and Figure \ref{fig:chaos-zoom} shows several levels of zoom on a six black hole system.

\vspace{-3mm}

\vspace{-6mm}

\section*{Summary \& Acknowledgements}

This work highlights a correspondence between light propagation in Majumdar-Papapetrou black hole spacetimes and classical optics in a variable refractive medium. By leveraging this analogy, we transform the challenge of relativistic ray tracing into a problem solvable with standard 3D graphics techniques, enabling efficient and accurate visualization of complex black hole configurations. I would like to thank Marcelo Seri for introducing me to the beautiful Majumdar Papapetrou solution, which started this project.

\vspace{-2mm}

    
{\setlength{\baselineskip}{13pt} 
\raggedright				
\bibliographystyle{bridges}
\bibliography{refs}
} 

\end{document}